\documentstyle[11pt,epsf,ssi]{article}

\newcommand{\beq}{\begin{equation}}
\newcommand{\eeq}{\end{equation}}
\newcommand{\beqa}{\begin{eqnarray}}
\newcommand{\eeqa}{\end{eqnarray}}

\newcommand{\NPB}[1]{{\it Nucl. Phys.}\ {\bf B{#1}}}
\newcommand{\PLB}[1]{{\it Phys. Lett.}\ {\bf B{#1}}}
\newcommand{\PRD}[1]{{\it Phys. Rev.}\ {\bf D{#1}}}

\newcommand{\lae}{\begin{array}{c}\,\sim\vspace{-21pt}\\< \end{array}}
\newcommand{\gae}{\begin{array}{c}\,\sim\vspace{-21pt}\\> \end{array}}
\newcommand{\semi}{;}

\def\slashchar#1{\setbox0=\hbox{$#1$}           % set a box for #1
   \dimen0=\wd0                                 % and get its size
   \setbox1=\hbox{/} \dimen1=\wd1               % get size of /
   \ifdim\dimen0>\dimen1                        % #1 is bigger
      \rlap{\hbox to \dimen0{\hfil/\hfil}}      % so center / in box
      #1                                        % and print #1
   \else                                        % / is bigger
      \rlap{\hbox to \dimen1{\hfil$#1$\hfil}}   % so center #1
      /                                         % and print /
   \fi}                                         %

\begin{document}

%\Draft

\begin{titlepage}
\def\thepage {}        % Kill page numbering
\title{
Dynamical Electroweak Symmetry Breaking \\
and the Top Quark\thanks{\rm An abbreviated version$^1$ of this
talk was presented at the {\it Workshop on Top Quark Physics}, Iowa
State University, Ames, IA, May 25-26, 1995 and the {\it Yukawa
International Seminar `95}, Yukawa Institute, Kyoto, Aug. 21-25,
1995.}
}

\author{
R. Sekhar Chivukula \\
Department of Physics, Boston University, \\
590 Commonwealth Ave., Boston MA 02215 \\
e-mail: sekhar@bu.edu}

\maketitle
\medskip
\centerline{\em Talk presented at SLAC Topical Workshop}
\centerline{\em Stanford, July 19-21, 1995}
\medskip
\centerline{\em BUHEP-95-23 \& hep-ph/9509384}

\begin{abstract}
In this talk I discuss theories of dynamical electroweak symmetry
breaking, with emphasis on the implications of a heavy top-quark
on the weak-interaction $\rho$ parameter.
\pagestyle{empty}
\end{abstract}
\bigskip

\end{titlepage}

%%%%%%%%%%%%%%%%%%%%%%%%%%%%%%%%

\section{What's Wrong with the Standard Model?}
\setcounter{equation}{0}

In the standard one-doublet Higgs model one introduces a fundamental
scalar doublet of $SU(2)_W$:
\beq
\phi=\left(\matrix{\phi^+ \cr \phi^0 \cr}\right)
{}~~~,
\eeq
which has a potential of the form
\beq
V(\phi)=\lambda \left(\phi^{\dagger}\phi - {v^2\over 2}\right)^2
{}~~~.
\eeq
In the potential, $v^2$ is assumed to be positive in order to
favor the generation of a non-zero vacuum expectation value for
$\phi$.  This vacuum expectation value breaks the electroweak
symmetry, giving mass to the $W$ and $Z$.

This explanation of electroweak symmetry breaking is unsatisfactory
for a number of reasons.  For one thing, this model does not give a
dynamical explanation of electroweak symmetry breaking. For another,
when embedded in theories with additional dynamics at higher energy
scales, these theories are technically unnatural \cite{thooft} .

Perhaps most unsatisfactory, however, is that theories of fundamental
scalars are probably ``trivial'' \cite{wilson}, {\it i.e.}, it is not possible
to construct an interacting
theory of scalars in four dimensions that is valid to arbitrarily
short distance scales.  In quantum field theories, fluctuations in the
vacuum screen charge -- the vacuum acts as a dielectric
medium. Therefore there is an effective coupling constant which
depends on the energy scale ($\mu$) at which it is measured. The
variation of the coupling with scale is summarized by the
$\beta$--function of the theory
\beq
\beta(\lambda) = \mu{d\lambda\over d\mu}
{}~~~.
\eeq
The only coupling in the Higgs sector of the standard model is the
Higgs self-coupling $\lambda$. In perturbation theory, the
$\beta$-function is calculated to be
\beq
{\lower15pt\hbox{\epsfysize=0.5 truein \epsfbox{fig4.eps}}}
\rightarrow \beta = {3\lambda^2 \over 2 \pi^2}
{}~~~.
\eeq
Using this $\beta$--function, one can compute the behavior of the
coupling constant as a function of the scale\footnote{Since these
expressions were computed in perturbation theory, they are only valid
when $\lambda(\mu)$ is sufficiently small. For large coupling we must
rely on non-perturbative lattice monte-carlo
studies\cite{kuti,neuberger}, which show behavior similar to that
implied by the perturbative expressions derived here.}. One finds that
the coupling at a scale $\mu$ is related to the coupling at some
higher scale $\Lambda$ by
\beq
{1\over\lambda(\mu)}={1\over\lambda(\Lambda)}
+{3\over 2\pi^2}\log{\Lambda\over\mu}
{}~~~.
\eeq
In order for the Higgs potential to be stable, $\lambda(\Lambda)$
has to be positive. This implies
\beq
{1\over\lambda(\mu)} \ge {3\over
2\pi^2}\log{\Lambda\over\mu}
{}~~~.
\eeq
Thus, we have the bound
\beq
\lambda(\mu)\le{2\pi^2\over 3\log\left({\Lambda\over\mu}\right)}
{}~~~.
\eeq
If this theory is to make sense to arbitrarily short distances, and
hence arbitrarily high energies, we should take $\Lambda$ to $\infty$
while holding $\mu$ fixed at about 1 TeV. In this limit we see
that the bound on $\lambda$ goes to zero. In the continuum limit, this
theory is trivial; it is free field theory.

The theory of a relatively light weakly coupled Higgs boson, can be
self-consistent to a very high energy.  For example, if the theory is
to make sense up to a typical GUT scale energy, $10^{16}$ GeV, then
the Higgs boson mass has to be less than about 170 GeV \cite{maiani}.  In
this sense, although a theory with a light Higgs boson does not really
answer any of the interesting questions ({\it e.g.}, it does not
explain {\it why} $SU(2)_W\times U(1)_Y$ breaking occurs), the theory does
manage to postpone the issue up to higher energies.

\section{Dynamical Electroweak Symmetry Breaking}
\subsection{Technicolor}

Technicolor\cite{weinberg} theories strive to explain electroweak
symmetry breaking in terms of physics operating at an energy scale of
order a TeV.  In technicolor theories, electroweak symmetry breaking
is the result of chiral symmetry breaking in an asymptotically-free,
strongly-interacting gauge theory with massless fermions.  Unlike
theories with fundamental scalars, these theories are technically
natural: just as the scale $\Lambda_{QCD}$ arises in QCD by
dimensional transmutation, so too does the weak scale $v$ in
technicolor theories.  Accordingly, it can be exponentially smaller
than the GUT or Planck scales.  Furthermore, asymptotically-free
non-abelian gauge theories may be fully consistent quantum field
theories.

In the simplest theory \cite{weinberg} one introduces doublet of new
massless fermions

\beq
\Psi_L=\left(
\begin{array}{c}
U\\D
\end{array}
\right) _L\,\,\,\,\,\,\,\,
U_R,D_R
\eeq

\noindent
which are $N$'s of a technicolor gauge group
$SU(N)_{TC}$. In the absence of electroweak interactions, the Lagrangian for
this theory may be written

\beqa
{\cal L} &=& \bar{U}_L i\slashchar{D} U_L+
\bar{U}_R i\slashchar{D} U_R+\\
 & &\bar{D}_L i\slashchar{D} D_L+
\bar{D}_R i\slashchar{D} D_R
\eeqa

\noindent
and thus has an $SU(2)_L\times SU(2)_R$ chiral symmetry. In analogy with QCD,
we
expect that when technicolor becomes strong,

\beq
\langle \bar U_LU_R\rangle
=\langle \bar D_LD_R\rangle \neq 0,
\eeq

\noindent
which breaks the global chiral symmetry group down to $SU(2)_{L+R}$, the
vector subgroup (analogous to isospin in QCD).

If we weakly gauge $SU(2) \times U(1)$, with the left-handed
technifermions forming a weak doublet and identify hypercharge with a
symmetry generated by a linear combination of the $T_3$ in $SU(2)_R$
and technifermion number, then chiral symmetry breaking will result in
the electroweak gauge group's breaking down to electromagnetism. The
Higgs mechanism then produces the appropriate masses for the $W$ and
$Z$ bosons if the $F$-constant of the technicolor theory (the analog
of $f_\pi$ in QCD) is approximately 246 GeV. (The residual
$SU(2)_{L+R}$ symmetry insures that, to lowest-order, $M_W = M_Z
\cos\theta_W$ and the weak interaction $\rho$-parameter equals one at
tree-level \cite{custodial}.)

\subsection{Top-Mode and Strong-ETC Models}

There is also a class of theories in which the scale ($M$) of the
dynamics responsible for (all or part of) electroweak symmetry
breaking can, in principle, take any value of order a TeV or greater.
These models, inspired by the Nambu--Jona-Lasinio (NJL) model
\cite{nambu} of chiral symmetry breaking in QCD, involve a strong, but
{\it spontaneously broken}, non-confining gauge interaction. Examples
include top quark condensate (and related) models
\cite{topmode,bardeen,fourgen,brokentc,models}, as well as models with
strong extended technicolor interactions \cite{strongETC}. When the
strength of the effective four-fermion interaction describing the
broken gauge interactions -- {\it i.e.} the strength of the extended
technicolor interactions in strong ETC models or the strength of other
gauge interactions in top-condensate models -- is adjusted close to
the critical value for chiral symmetry breaking, the high-energy
dynamics may play a role in electroweak symmetry breaking without
driving the electroweak scale to a value of order $M$.

The high-energy dynamics must have the appropriate properties in order
for it to play a role in electroweak symmetry breaking \cite{cohen}: If the
coupling constants of the high-energy theory are small, only
low-energy dynamics (such as technicolor) can contribute to
electroweak symmetry breaking. If the coupling constants of the
high-energy theory are large and the interactions are attractive in
the appropriate channels, chiral symmetry will be broken by the
high-energy interactions and the scale of electroweak symmetry
breaking will be of order $M$.  If the transition between these two
extremes is continuous, {\it i.e.} if the chiral symmetry breaking
phase transition is {\it second order} in the high-energy couplings,
then it is possible to adjust the high-energy parameters so that the
dynamics at scale $M$ can contribute to electroweak symmetry breaking.
The adjustment of the high-energy couplings is a reflection of the
fine-tuning required to create a hierarchy of scales.

What is crucial is that the transition be (at least approximately)
second order in the high-energy couplings.  If the transition is first
order, then as one adjusts the high-energy couplings the scale of
chiral symmetry breaking will jump discontinuously from approximately
zero at weak coupling to approximately $M$ at strong coupling.
Therefore, if the transition is first order, it will generally not be
possible to maintain any hierarchy between the scale of electroweak
symmetry breaking and the scale of the high-energy dynamics.

If the transition is second order and if there is a {\it large}
hierarchy of scales ($M \gg$ 1 TeV), then close to the transition the
theory may be described in terms of a low-energy effective Lagrangian
with composite ``Higgs'' scalars -- the Ginsburg-Landau theory of the
chiral phase transition.  However, if there is a large hierarchy, the
arguments of triviality given in the first section apply to the
effective low-energy Ginsburg-Landau theory describing the composite
scalars: the effective low-energy theory would be one which describes
a {\it weakly} coupled theory of (almost) fundamental scalars, despite
the fact that the ``fundamental'' interactions are strongly
self-coupled!

%%%%%%%%%%%%%%%%%%%%%%%%%%%%%%%%%%%%
\section{$m_t$ in Models of Dynamical EWSB}
\label{sec:mt}
\setcounter{equation}{0}

In technicolor models, the masses of the ordinary fermions are due to
their coupling to the technifermions, whose chiral-symmetry breaking
is responsible for electroweak symmetry breaking. This is
conventionally \cite{ETC} assumed to be due to additional, broken,
extended-technicolor (ETC) gauge-interactions:

\beq
\epsfxsize 5cm \centerline{\epsffile{tETC.eps}}
\eeq

\noindent
which leads to a mass for the top-quark

\beq
m_t \approx {g^2 \over M^2_{ETC}} \langle\bar{U} U \rangle_{M_{ETC}}\ ,
\eeq

\noindent
where we have been careful to note that it is the value of the
technifermion condensate renormalized at the scale $M_{ETC}$ which is
relevant.

For a QCD-like technicolor, there is no substantial
difference between $\langle\bar{U} U \rangle_{M_{ETC}}$ and
$\langle\bar{U} U \rangle_{\Lambda_{TC}}$, and we can use naive
dimensional analysis \cite{dimanal} to estimate the technifermion
condensate, arriving at a top-quark mass

\beq
m_t \approx {g^2 \over M^2_{ETC}} 4 \pi F^3\ ,
\eeq

\noindent
We can invert this relation to find the characteristic mass-scale of
top-quark mass-generation

\beq
{M_{ETC} \over g} \approx 1\ {\rm TeV} \left({F  \over 246\ {\rm
GeV}}\right)^{3\over 2} \left({175\ {\rm GeV} \over m_t}\right)^{1\over
2}.
\label{blitz}
\eeq

We immediately see that the scale of top-quark mass generation is
likely to be {\it quite} low, unless the value of the technifermion
condensate ($\langle\bar{U} U \rangle_{M_{ETC}}$) can be raised
significantly above the value predicted by naive dimensional
analysis. The prospect of such a low ETC-scale is both tantalizing and
problematic. As we will see in the next section, constraints from the
deviation of the weak interaction $\rho$ parameter from one suggest
that the scale may have to be larger than one TeV.

There have been two approaches to enhance the technifermion condensate
which have been discussed in the literature: ``walking''
\cite{walking} and ``strong-ETC'' \cite{strongETC}. In a walking
theory, one arranges for the technicolor coupling constant to be
approximately constant and large over some range of momenta. The
maximum enhancement that one might expect in this scenario is

\beq
\langle\bar{U} U \rangle_{M_{ETC}} \approx
\langle\bar{U} U \rangle_{\Lambda_{TC}} \left({M_{ETC} \over
\Lambda_{TC}}\right)^{\gamma(\alpha_{TC})}\ ,
\eeq

\noindent
where $\gamma(\alpha_{TC})$ is the anomalous dimension of the
technifermion mass operator (which is possibly as large as one). As
described above, however, we expect that $M_{ETC}$ cannot be too much
higher than $\Lambda_{TC}$, and therefore that the enhancement due to
walking is not sufficient to reconcile the top-quark mass and an ETC
scale higher than a TeV.

\begin{figure}[htb]
\vspace{0.5cm}
\epsfxsize 8cm \centerline{\epsffile{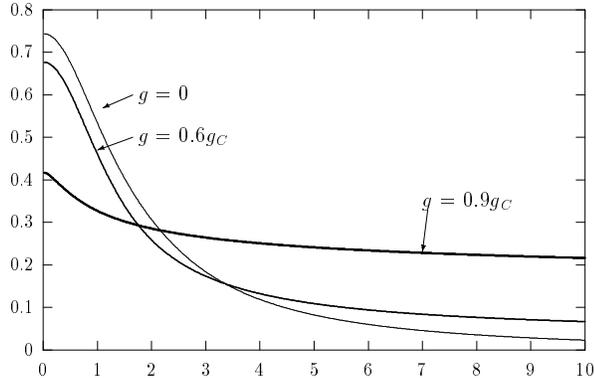}}
\caption{Plot\protect\cite{evans} of technifermion self energy
vs. momentum (both measured in TeV), as predicted by the gap-equation
in the rainbow approximation, for various strengths of the ETC
coupling relative to their critical value $g_C$.}
\label{figone}
\vspace{0.5cm}
\end{figure}

The strong-ETC alternative is potentially more promising. As the size
of the ETC-coupling at the ETC scale approaches the critical value for
chiral symmetry breaking, it is possible to enhance the running
technifermion self-energy $\Sigma(k)$ at large momenta (see
Fig. \ref{figone}).  Since the technifermion condensate is related to
the trace of the fermion propagator.

\beq
\langle\bar{U} U \rangle_{M_{ETC}} \propto \int^{M^2_{ETC}}
dk^2 \Sigma(k)\ ,
\eeq

\noindent
a slowly-falling running-mass translates to an enhanced
condensate\footnote{More physically, in terms of the relevant
low-energy theory, it can be shown that the enhancement of the
top-quark mass is due to the dynamical generation of a light scalar
state \cite{cohen,Comp}}.

\begin{figure}[htb]
\vspace{0.5cm}
\epsfxsize 8cm \centerline{\epsffile{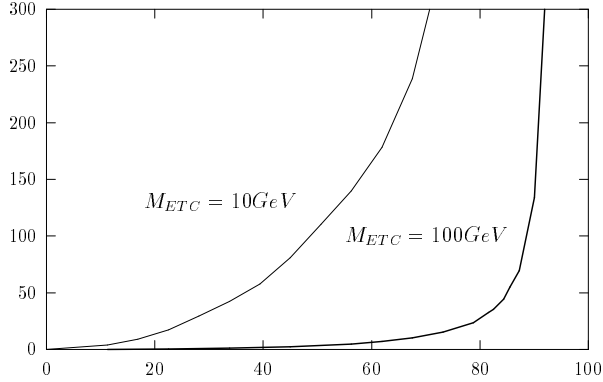}}
\caption{ Plot\protect\cite{evans} of top mass (in GeV) vs. ETC
coupling ($g/g_C$ in \%), as predicted by gap-equation in the rainbow
approximation, for ETC scales of 10 and 100 TeV.}
\label{figtwo}
\vspace{0.5cm}
\end{figure}

Unfortunately, there is no such thing as a free lunch.  As we see from
Fig. \ref{figtwo}, the enhancement of the technifermion self-energy in
strong-ETC theories comes at the cost of a ``fine-tuning" of the
strength of the ETC coupling relative to the critical value where the
ETC interactions would, in and of themselves, generate chiral symmetry
breaking. In the context of the NJL approximation, we find that
enhancement of the top quark mass is directly related to the severity
of this adjustment. In particular, if we denote the critical value of
the ETC coupling by $g_C$, in the NJL approximation we find \cite{nambu}

\beq
{\langle\bar{U} U \rangle_{\Lambda_{TC}}\over \langle\bar{U} U
\rangle_{M_{ETC}}} \approx {\Delta g^2 \over g^2_c}
\label{price}
\eeq

\noindent
where $\Delta g^2 \equiv g^2 - g^2_C$.

%%%%%%%%%%%%%%%%%%%%%%%%%%%%%%%%%%%%
\section{$\Delta\rho_*$}
\label{sec:deltarho}
\setcounter{equation}{0}

The physics which is responsible for top-quark mass generation must
violate custodial $SU(2)$ since, after all, this physics must give
rise to the disparate top- and bottom-quark masses. The danger is that
this isospin violation will ``leak'' into the $W$ and $Z$ gauge-boson
masses and to give rise to a deviation of the weak interaction
$\rho$-parameter from one.

\subsection{Direct Contributions}

As emphasized by Appelquist, Bowick, Cohler, and Hauser
\cite{isospin2}, ETC operators which violate custodial isospin by two
units ($\Delta I = 2$) are particularly dangerous. Denoting the
right-handed technifermion doublet by $\Psi_R$, consider the operator

\beq
{g^2 \over M^2} \left(\bar{\Psi}_R \gamma_\mu \sigma_3 \Psi_R\right)^2~~,
\label{cohler}
\eeq

\noindent
which can result in the (mass-)mixing of the $Z$ with an isosinglet ETC
gauge-boson

\beq
\epsfxsize 8cm \centerline{\epsffile{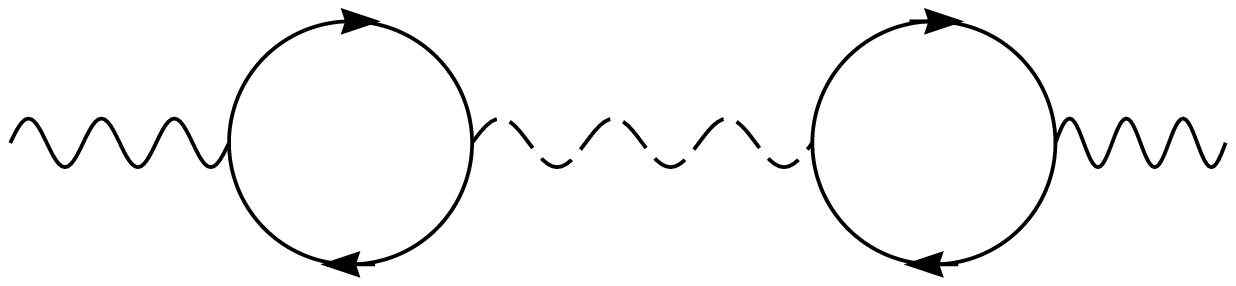}}
\eeq

\noindent
and hence a contribution to $\Delta\rho$.
Contributions of this sort
arise naturally in ETC-models which give rise to the
top-quark mass\cite{wu}.

If there are $N_D$ doublets of the technifermions $\Psi$, and they
give rise to a contribution to $M^2_W$ proportional to $N_D F^2$, the
contribution of the operator in eqn. (\ref{cohler}) to the $\rho$
parameter can be estimated to be

\beqa
\Delta\rho_* & \approx &  {2 g^2 \over M^2} {N^2_D F^4 \over v^2} \\
& \approx &  12\%\  g^2 \left({N_D F^2 \over (246 \ {\rm
GeV})^2}\right)^2 \left({1 \ {\rm TeV} \over M}\right)^2 ~.
\label{important}
\eeqa

\begin{figure}[htb]
\vspace{0.5cm}
\epsfxsize 10cm \centerline{\epsffile{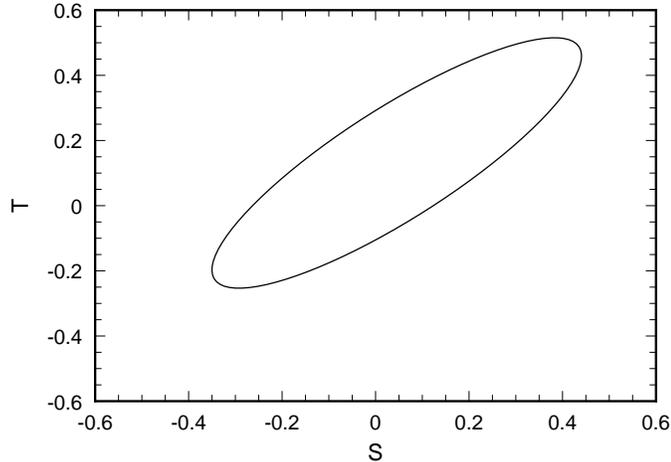}}
\caption{The ellipse\protect\cite{tc2isospin} in the $S-T$ plane which projects
onto the 95\% confidence range for $T$. Note that $\Delta \rho_* =
\alpha T$.}
\label{figthree}
\vspace{0.5cm}
\end{figure}

\noindent
Current limits (see Fig. \ref{figthree}) on the parameter $T$ ($\Delta
\rho_* = \alpha T$) imply that $\Delta\rho_* \lae 0.4\% $.

There are two ways\footnote{It is also conceivable
\cite{revenge} that there are additional
isospin-asymmetric contributions --- say, from relatively light
pseudo-Goldstone bosons --- which give rise to {\em negative}
contributions to $T$ and cancel some or all of the positive
contributions discussed here.} in which one may try to satisfy this
constraint.  The equation above implies

\beq
{M\over g} \gae  5.5\ {\rm TeV} \left( {N_D F^2 \over (246\ {\rm
GeV})^2}\right)\ .
\eeq

\noindent
If $N_D F^2 \approx (246\ {\rm GeV})^2$, that is if the sector giving
rise to the top-quark mass is responsible for the bulk of EWSB, then
the scale $M$ must be much larger than the naive 1 TeV expectation in
QCD-like technicolor.  Comparing this with eqns. (\ref{blitz}) and
(\ref{price}) above, we see that the enhancement of the condensate
needed requires a fine-tuning of order 3\%
($\approx (1/5.5)^2$) in
order to produce a top-quark mass of order 175 GeV.

Alternatively, we may re-write the bound as

\beq
F \lae {105\ {\rm GeV} \over \sqrt{N_D}} \left( {M/g \over 1\ {\rm
TeV}}\right)^{1\over 2}
\eeq

\noindent
If $M/g$ is of order 1 TeV, it is necessary that the sector
responsible for top quark mass generation {\em not} give rise to the
bulk of EWSB.  While this case is counter-intuitive (after all, the
third generation is the heaviest!), it may in fact provide a
resolution to the issue of how large isospin breaking can exist in the
fermion mass spectrum without leaking into the $W$ and $Z$ masses.
This is essentially what happens in multiscale models
\cite{multi,chiraltc} and in top-color assisted technicolor
\cite{top-color}.  Such hierarchies of technifermion masses are also
useful for reducing the predicted value of $S$ in technicolor
models\footnote{Recently the experimental upper bound on $S$ has been
relaxed, so that positive values of $S$ are allowed ( $S
< 0.4$ at the 95\% confidence level) \cite{tc2isospin}. }
\cite{revenge}.

\subsection{Indirect Contributions}

A second class of potentially dangerous contributions come from
isospin violation in the technifermion mass spectra.  In a manner
analogous to the contribution\cite{Drho} of the $t-b$ mass splitting
to $\Delta \rho$, any difference in the dynamical masses of two
technifermions in the {\em same} doublet will give rise to deviations
in the $\rho$ parameter from one. The size of this effect can be
estimated a la Pagels-Stokar \cite{pagels}. Using this approximation,
we find that the contributions to the loop diagram from low-momenta
dominate and

\beq
\Delta \rho_* \propto {N_D\, d \over 16 \pi^2}
\left({\Sigma_U(0) - \Sigma_D(0)}\over v\right)^2
\eeq

\noindent
where $N_D$ and $d$ are the the number of doublets and dimension of
the technicolor representation respectively.  Since we require $\Delta
\rho_* \lae 0.4$\%, the equation above implies

\beq
N_D\, d \left({\Delta\Sigma(0)\over m_t}\right)^2 \lae 1.3 \, \, .
\label{ilimit}
\eeq

\noindent
{}From this we see that, $\Delta \Sigma(0)$ must be less than of order
$m_t$ (perhaps, given the crude approximations involved, one may be
able to live with $d=2$ in the fundamental of and $SU(2)$ technicolor
group with one doublet).

However, if the $t$ and $b$ get their mass from the same technidoublet, then at
the ETC-scale we expect that there is no difference between the $t$,
$b$ and the corresponding technifermions \cite{strongETC}

\beqa
\Delta\Sigma(M_{ETC})& \equiv & \Sigma_U(M_{ETC}) - \Sigma_D(M_{ETC})
\approx  \nonumber \\
\Delta m(M_{ETC}) & \equiv & m_t(M_{ETC}) - m_b(M_{ETC})~.
\label{problem}
\eeqa

\noindent
Furthermore, if QCD is the only interaction which contributes to the
scaling of the $t$ and $b$ masses, we expect $\Delta m(M_{ETC}) \approx
m^{pole}_t$, and from scaling properties of the technifermion
self-energies, we expect $\Delta\Sigma(0) \gae \Delta
\Sigma(M_{ETC})$.

There are two ways to avoid these constraints.  One is that perhaps
there are {\it additional} interactions which contribute to the
scaling of the top- and bottom-masses below the ETC scale, and hence
that $\Delta m(M_{ETC}) \ll m_t^{pole}$. This would be the case if the
$t$ and/or $b$ get only a {\em portion} of their mass from the
technicolor interactions, and would imply that the third generation
must have (strong) interactions different from the technifermions (and
possibly from the first and second generations).  Another possibility
is that the $t$ and $b$ get mass from {\em different} technidoublets,
each of which have isospin-symmetric masses. The first alternative is
the solution chosen in top-color assisted technicolor models (see
below), while the latter has only recently begun to be
explored \cite{isosymmetric}.

%%%%%%%%%%%%%%%%%%%%%%%%%%%%%%%%
\section{Case Study: Top-Color Assisted Technicolor}
\label{sec:tcii}
\setcounter{equation}{0}

Recently, Hill has combined aspects of two different approaches to
dynamical electroweak symmetry breaking into a model which he refers
to as top-color assisted technicolor \cite{top-color}.
In this model a top-condensate is driven by the combination of a
strong, but spontaneously broken and non-confining, isospin-symmetric
top-color interaction and an additional (either weak or strong)
isospin-breaking $U(1)$ interaction which couple only to the third
generation quarks.

At low-energies, the top-color and hypercharge interactions of the
third generation quarks may be approximated by four-fermion operators
\cite{top-color}.

\beq
{\cal L}_{4f} = -{{4\pi
\kappa_{tc}}\over{M^2}}\left[\overline{\psi}\gamma_\mu {{\lambda^a}\over{2}}
\psi \right]^2
-{{4\pi \kappa_1}\over{M^2}}\left[{1\over3}\overline{\psi_L}\gamma_\mu  \psi_L
+{4\over3}\overline{t_R}\gamma_\mu  t_R
-{2\over3}\overline{b_R}\gamma_\mu  b_R
\right]^2~,
\label{L4t}
\eeq

\noindent
where $\psi$ represents the top-bottom doublet, $\kappa_{tc}$ and
$\kappa_1$ are related respectively to the top-color and $U(1)$
gauge-couplings squared, and where (for convenience) we have assumed
that the top-color and $U(1)$ gauge-boson masses are comparable and of
order $M$. The first term in equation (\ref{L4t}) arises from the
exchange of top-color gauge bosons, while the second term arises from
the exchange of the new $U(1)$ hypercharge gauge boson which has
couplings proportional to the ordinary hypercharge couplings.  In
order to produce a large top quark mass without giving rise to a
correspondingly large bottom quark mass, the combination of the
top-color and extra hypercharge interactions are assumed to be
critical in the case of the top quark but not the bottom quark.  The
criticality condition for top quark condensation in this model is
then:
\beq
\kappa^t_{eff} = \kappa_{tc} +{1\over3}\kappa_1 >
\kappa_c = {{3\pi}\over{8}} >
\kappa^b_{eff}=\kappa_{tc} -{1\over 6}\kappa_1~.
\label{kc}
\eeq

The contribution of the top-color sector to electroweak symmetry
breaking can be quantified by the F-constant of this sector.  In the
NJL approximation \cite{nambu}, for $M$ of order 1 TeV,
and $m_t \approx 175$ GeV, we find
\beq
f_t^2 \equiv    {{N_c }\over{8\pi^2}}\, m_t^2
\log\left({{M^2}\over{m_t^2}}\right) \approx (64\ {\rm GeV})^2~.
\label{ft}
\eeq
As $f_t$ is small compared to 246 GeV, there must be additional
dynamics which is largely responsible for giving rise to the $W$ and
$Z$ masses. In top-color assisted technicolor, technicolor
interactions play that role.

\subsection{Direct Isospin Violation}

Technifermions are necessary to produce the bulk of EWSB and to give
mass to the light fermions. However, the heavy and light fermions must
mix --- hence, we would naturally expect that at least some of the
{\it technifermions} carry the extra $U(1)$ interaction.  If the
additional $U(1)$ interactions violate custodial symmetry\footnote{It
has been noted \cite{isosymmetric} that if the top- and bottom-quarks
receive their masses from {\em different} technidoublets, it is
possible to assign the extra $U(1)$ quantum numbers in a custodially
invariant fashion.}, the $U(1)$ coupling will have to be quite small
to keep this contribution to $\Delta \rho_*$ small \cite{tc2isospin}.
We will illustrate this in the one-family technicolor \cite{onefam}
model, assuming that techniquarks and technileptons carry
$U(1)$-charges proportional to the hypercharge of the corresponding
ordinary fermion\footnote{Note that this choice is anomaly-free.}. We
can rewrite the effective $U(1)$ interaction of the technifermions as

\beq
{\cal L}_{4T1} =-{{4\pi
\kappa_1}\over{M^2}}\left[{1\over3}\overline{\Psi}\gamma_\mu  \Psi
+\overline{\Psi}_R\gamma_\mu  \sigma^3 \Psi_R
-\overline{L} \gamma_\mu L
+ \overline{L}_R \gamma_\mu\sigma^3 L_R
\right]^2~,
\label{L4T}
\eeq

\noindent
where $\Psi$ and $L$ are the techniquark and technilepton
doublets respectively.

{}From the analysis given above (eqn. (\ref{important})), we see that
the contribution to $\Delta
\rho_*$ from degenerate technifermions is \cite{tc2isospin}:
\beq
\Delta \rho_{*}^{\rm T} \approx 152\% \
\kappa_1 \left({{1\ {\rm TeV}}\over{M}}\right)^2~.
\label{rhoT}
\eeq

\noindent
Therefore, if $M$ is of order 1 TeV and the extra $U(1)$ has
isospin-violating couplings to technifermions, $\kappa_1$ must be
extremely small.

\subsection{Indirect Isospin Violation}

In principle, since the isospin-splitting of the top and bottom are
driven by the combination of top-color and the extra $U(1)$, the
technifermions can be degenerate. In this case, the only indirect
contribution to the $\rho$ parameter at one-loop is the usual
contribution coming from loops of top- and bottom-quarks \cite{Drho}.
However, since there are additional interactions felt by the
third-generation of quarks, there are ``two-loop'' contributions of
the form

\beq
\epsfxsize 8cm \centerline{\epsffile{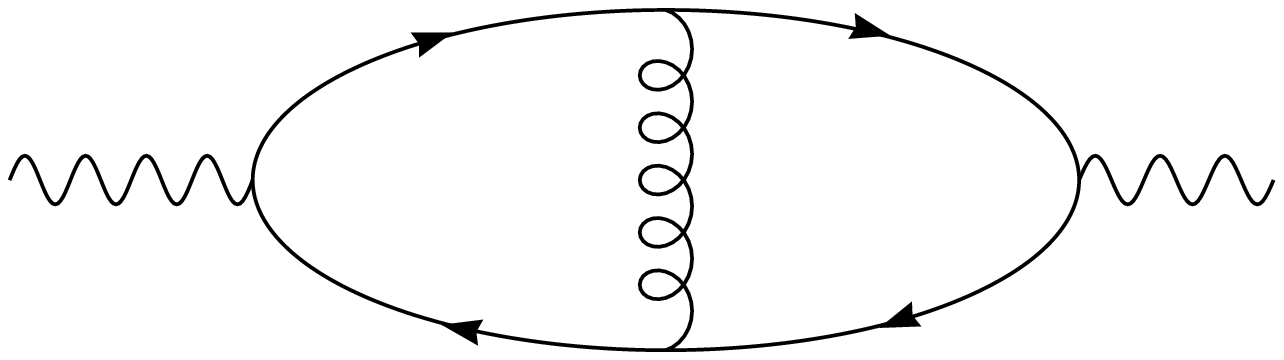}}
\label{twoloop}
\eeq
This contribution yields \cite{tc2isospin}
\beq
\Delta \rho_{*}^{\rm tc} \approx 0.53\%
\left({\kappa_{tc}\over \kappa_c}\right)
\left(1\ {\rm TeV}\over M\right)^2
\left(f_t \over 64\ {\rm GeV}\right)^4
{}~.
\label{rhot}
\eeq

\noindent
Combining this with eqn. (\ref{ft}), we find that
\beq
M \gae 1.4\ {\rm TeV}
\eeq
for $\kappa_{tc} \approx \kappa_c$.

\noindent
This immediately puts a constraint on the mass of the top-color gluon
which is comparable to the direct limits currently obtained by CDF
\cite{cdftwojet}.

\subsection{Fine-Tuning}

Finally, we must require that the sum of the effects of eqns. (\ref{rhoT})
and (\ref{rhot}) do not give rise to an experimentally disallowed
contribution to the $\rho$ parameter.  Equation (\ref{rhoT}) implies
that $\kappa_1$ must either be very small, or $M$ very large. However,
we must also simultaneously satisfy the constraint of eqn. (\ref{kc}),
which implies that

\beq
{\Delta\kappa_{tc}\over\kappa_c} = \left|{{\kappa_{tc}-\kappa_c}
\over \kappa_c}\right| \le
{1 \over 3} {\kappa_1\over \kappa_c}~,
\label{dktc}
\eeq

\begin{figure}[htb]
\vspace{0.5cm}
\epsfxsize 10cm \centerline{\epsffile{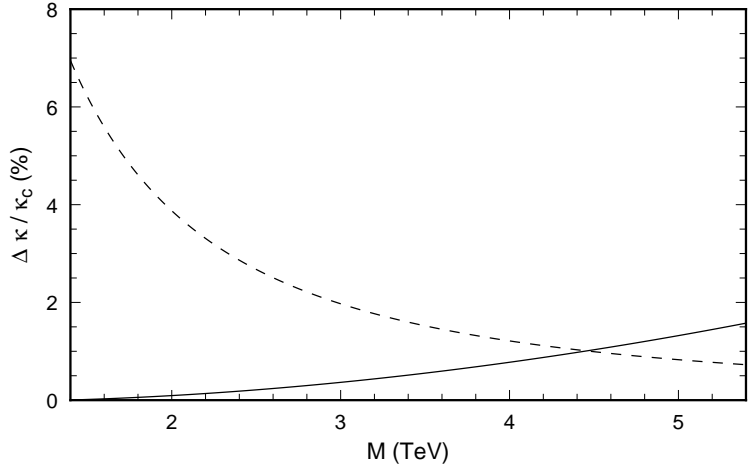}}
\caption{The amount of fine-tuning required\protect\cite{tc2isospin}
in the ${\rm TC}^2$ model.  The dashed line is the amount of
fine-tuning in $\Delta\kappa_{eff}$ required to keep $m_t$ much
lighter than $M$, see equation (\ref{dkefft}). The solid curve shows
the amount of fine-tuning (see equation (\ref{dktc})) in
$\Delta\kappa_{tc}$ required to satisfy the bound $\Delta \rho_* <
0.4$\%.  The region excluded by the experimental constraint on $\Delta
\rho_*$ is above the solid curve.}
\label{figfour}
\vspace{0.5cm}
\end{figure}

\noindent
Therefore, if $M$ is low and $\kappa_1$ is small, the
top-color coupling must be tuned close to the critical value for
chiral symmetry breaking. On the other hand, if $\kappa_1$ is not
small and $M$ is relatively large the {\em total} coupling of the
top-quark must be tuned close to the critical NJL value for
chiral symmetry breaking in order to keep the top-quark mass low.
The gap-equation for the Nambu--Jona-Lasinio model implies that
\beq
{\Delta\kappa_{eff}\over\kappa_c} =
{{\kappa^t_{eff}-\kappa_c}\over \kappa_c} =
{{{m^2_t\over M^2}\log{M^2\over m^2_t}} \over {1-{{m^2_t\over
M^2}\log{M^2\over m^2_t}}}}~.
\label{dkefft}
\eeq
These two constraints are shown in Fig. \ref{figfour}.  For $M>$ 1.4
TeV, we find that either $\Delta\kappa_{tc}/
\kappa_c$ or $\Delta\kappa_{eff}/\kappa_c$ must be tuned to
less than 1\%.  This trade-off in fine tunings is displayed in
figure 4.  For the ``best" case where both tunings are of order 1\%,
$M=4.5$ TeV.

%%%%%%%%%%%%%%%%%%%%%%%%%%%%%%
\section{Conclusions}
\label{sec:concl}

We have seen that a large top quark mass has a number of
important implications for dynamical electroweak symmetry
breaking:

\begin{itemize}
\item[$\bullet$]
A large top-quark mass naturally implies, in models of dynamical
electroweak symmetry breaking, the possibility of a correspondingly
low scale for the scale of top flavor-physics. While I have emphasized
the constraints on such physics arising from potential contributions
to the weak interaction $\rho$ parameter, there are also significant
constraints arising from the size of the $Z \to b\bar{b}$ branching
ratio\cite{zbb,wu}, as well as from contributions to $b \to s
\gamma$ and $B-\bar{B}$ mixing\cite{randall,balaji,kominis}.

\item[$\bullet$]
The physics responsible for the large isospin breaking in the $t-b$
mass splitting can lead to potentially dangerous ``direct" and ``indirect"
effects in the $W$ and $Z$ masses.

\item[$\bullet$]
The direct and indirect effects can be mitigated if the sector which
is responsible for the top- and bottom-masses does {\em not} provide the
bulk of electroweak symmetry breaking and, conversely, if the sector
responsible for the $W$ and $Z$ masses gives rise to only a {\em small
portion} of the top- and bottom-masses. This can happen only if the top
and bottom feel {\em strong} interactions which are not shared by the
technifermions and, possibly, the first two generations.

\item[$\bullet$]
In top-color assisted technicolor, the extra top-color interactions
give rise to additional indirect contributions to $\Delta \rho$, and
we must require that $M_{g} \gae 1.4\  {\rm TeV}$. Furthermore, If the extra
$U(1)$ has isospin-violating couplings to technifermions, we require
fine-tuning of order 1\%.

\end{itemize}

%%%%%%%%%%%%%%%%%%%%%%%%%%%%%%
\vfill\eject
\centerline{\bf Acknowledgments}

I thank Tom Appelquist, Nick Evans, and Ken Lane for helpful
conversations, Mike Dugan for help in preparing the manuscript, and
John Terning and Bogdan Dobrescu for collaboration\cite{tc2isospin} on
some of the work reported in this talk.  I also acknowledge the
support of an NSF Presidential Young Investigator Award and a DOE
Outstanding Junior Investigator Award.  {\em This work was supported
in part by the National Science Foundation under grant PHY-9057173,
and by the Department of Energy under grant DE-FG02-91ER40676.}

%%%%%%%%%%%%%%%%%%%%%%%%%%%%%%%%%%%%%%%%%%%%%%%%%%%%%%%%%%%%%%%%%%%%%

\end{document}